# Detection of Anomalies and Faults in Industrial IoT Systems by Data Mining: Study of CHRIST Osmotron Water Purification System

Mohammad Sadegh Sadeghi Garmaroodi, Faezeh Farivar, *Senior Member, IEEE*, Mohammad Sayad Haghighi, *Senior Member, IEEE*, Mahdi Aliyari Shoorehdeli, *Senior Member, IEEE*, Alireza Jolfaei, *Senior Member, IEEE*

***Abstract*—Industry 4.0 will make manufacturing processes smarter but this smartness requires more environmental awareness, which in case of Industrial Internet of Things, is realized by the help of sensors. This article is about industrial pharmaceutical systems and more specifically, water purification systems. Purified water which has certain conductivity is an important ingredient in many pharmaceutical products. Almost every pharmaceutical company has a water purifying unit as a part of its interdependent systems. Early detection of faults right at the edge can significantly decrease maintenance costs and improve safety and output quality, and as a result, lead to the production of better medicines. In this paper, with the help of a few sensors and data mining approaches, an anomaly detection system is built for CHRIST Osmotron® water purifier. This is a practical research with real-world data collected from SinaDarou Labs Co. Data collection was done by using six sensors over two-week intervals before and after system overhaul. This gave us normal and faulty operation samples. Given the data, we propose two anomaly detection approaches to build up our edge fault detection system. The first approach is based on supervised learning and data mining e.g. by support vector machines. However, since we cannot collect all possible faults data, an anomaly detection approach is proposed based on normal system identification which models the system components by artificial neural networks. Extensive experiments are conducted with the dataset generated in this study to show the accuracy of the data-driven and model-based anomaly detection methods.**

## I. Introduction

WATER is one of the most important raw materials in pharmaceutical drug manufacturing process. Control and monitoring of water purification process, including production, storage, distribution and microbiological and chemical quality checking is taken very seriously in pharmaceutical companies. Generally speaking, in dependable automated systems, a quick and precise fault detection and identification mechanism at the edge is essential for equipment reliability, reduction in the number of off-products as well as maintenance costs. Timely detection of anomalies or faults can also prevent failures cascade that might happen due to the interconnections between industrial systems.

In water purification systems, anomalies and faults may be caused by over-voltage, over-current, overloading in high-pressure pumps, failure or malfunctioning of PLC cards and sensors, pipes leakage, sedimentation in pipes, etc. Since any malfunction in medicine-related systems can potentially be hazardous and affect public health, irregularities are taken seriously. Purified water which has certain conductivity is an important ingredient in many pharmaceutical products. Malfunctioning of water purification systems can result in poor water quality, reduced output or even plant shutdown.

There are many non-expensive and non-intrusive techniques for anomaly detection which are classified into model-based and data-driven groups [1, 2]. Model-based approaches work based on the model of physical systems, and can be further divided into observer-based and system identification-based subgroups [3]. Modeling of industrial processes is a critical issue because most real-world industrial systems are nonlinear, multi-input and multi-output, and due to interrelated parameters and noises, they have un-modelled dynamics and uncertainty [4]. If the basic model of a process is not well represented, the fault detection or diagnostic system will not be sufficiently robust in the presence of noise and disturbance. This implies that a model-driven fault/anomaly detection mechanism suits the cases in which the process is simple, low dimensional or can be accurately described by mathematical relations [5, 6].

When the system behavior is unknown and too complex, or its model is incomplete and unavailable, applying data-driven methods are prescribed. There are two main approaches for data driven fault/anomaly detection; unsupervised and supervised learning [7].

This paper develops two anomaly detection approaches in industrial IoT and more specifically, for water purification systems, both of which are based on machine learning and data mining. The proposed methods are tested on a CHRIST Osmotron® water purification system which is widely used in

M. Sadeghi is with the Department of Computer and Mechatronics Engineering, Science and Research Branch, Islamic Azad University, Tehran, Iran, Email: sadeghi@anslab.org.

F. Farivar (corresponding author) is with the Department of Computer and Mechatronics Engineering, Science and Research Branch, Islamic Azad University, Tehran, Iran, Email: f.farivar@srbiau.ac.ir, farivar@ieee.org.

M. Sayad Haghighi is with the School of Electrical and Computer Engineering, College of Engineering, University of Tehran, Iran, Email: sayad@ut.ac.ir, sayad@ieee.org.

M. Aliyari is with the Department of Mechatronics, Faculty of Electrical Engineering, K. N. Toosi University of Technology, Tehran, Iran, Email: aliyari@kntu.ac.ir.

A. Jolfaei is with the Department of Computing, Macquarie University, Australia, Email: alireza.jolfaei@mq.edu.au.

the the production process of pharmaceutical products. This system works based on a combination of reverse osmosis and electro-deionization and consists of three parts; Osmotron, cold water loop and hot water loop. Generally, by passing potable water through hardening columns, membranes, and electro-deionization in Osmotron unit, the purified water enters the cold and hot water loop for distribution and consumption. The conductivity of Osmotron output water has to be below one micro-siemens. More details on the system will be given in the subsequent sections,

In this study, data collection was the first challenge. To tackle this problem, six industrial sensors as well as one data recorder were installed on a CHRIST Osmotron system at SinaDarou Lab Co. to record and store sensors' information during two intervals, before and after the system was overhauled. Data were pre-processed and then fed into the anomaly detection systems. In the first approach, the collected data were divided into two classes by using support vector machines. In the second approach, the normal system was identified by an artificial neural network. Then, anomalies were detected based on the residual signal with static and adaptive thresholds. Experiments demonstrated the effectiveness of both strategies though the second one is more generic for anomaly detection.

The main contributions of this paper are as follows:

- Real-world data collection from a CHRIST Osmotron® water purification system in faulty and normal conditions.
- Designing data-driven and model-driven edge anomaly/fault detectors for industrial water purification systems based on data mining and machine learning principles.

The rest of this paper is organized as follows. In Section II, we briefly review related studies. Section III introduces the process and CHRIST Osmotron unit for water purification. In Section IV, machine learning-based approaches are developed for anomaly detection. Results are presented in Section V. Finally, the paper is concluded in Section VI.

## II. Related Work

Traditional methods for anomaly / fault detection are based on the random inspection of a system or a machine. For instance, overheating of an electric motor can be diagnosed as a fault. But the inefficiency of people, longtime high-intensity work, and random inspection make the traditional methods unreliable [8]. While nowadays diversity data-driven anomaly detection techniques have been developed and proposed for automatic control systems. These methods can discover meaningful rules to represent the information among the system variables only using time series data. Recently, anomaly detection is one of active research topics to increasing the safety and reliability of dynamical systems.

Because of the importance of water in pharmaceutical drug manufacture and water supply companies, several studies pay attention to the issue of anomaly detection using machine learning methods. In [9], an anomaly detection algorithm is designed to check the water quality by dual time-moving windows. The algorithm works based on autoregressive linear model. It has been tested on PH water quality data. In [10], SVM and artificial neural networks are used for water qualification. In [11], least squares support vector machine (LS-SVM) and particle swarm optimization methods have been applied to predict the water quality. This paper applied recurrent neural networks as a dynamical System. Moreover, they modelled the problem with multiple input variables, this can be useful in time series forecasting against to classical linear methods that the adaptation to multivariate or multiple input forecasting problems is difficult. Paper [12] presented a survey of machine learning methods to anomaly detection on the quality data of drinking-water. According to their achievements, the review encompasses both traditional ML and deep learning (DL) approaches. Having fair comparisons between published studies is difficult because of the difference in their data-set, models, parameters, etc. It is suggested to apply deep learning because of its advantages in feature learning accuracy. In [13], anomaly detection of the water quality data is studied using logistic regression, linear discriminant analysis, SVM, artificial neural network, deep neural network, recurrent neural network, and long short-term memory. The applied methods are compared using F-Score metric. In [14], machine learning and statistic learning approaches are studied to solve the problem of anomaly detection for water quality data. Different machine and static learning methods for time series like local outlier factor, isolation forest, robust random cut forest, seasonal hybrid extreme studentized deviate, and exponential moving average are applied to anomaly detection in a wastewater treatment plant. Comparison results are presented to depict the power of the methods. Generally, it is obviously that machine learning methods are appropriate tools to solve anomaly detection.

Nowadays, studies on Internet of Things (IoT), Cyber physical Systems (CPS), and Networked Control Systems (NCS) are significantly increased. Hence, anomaly detection algorithms have been paid much attention to in the mentioned systems [15, 16]. Reference [17] classified anomaly detection algorithms to four groups: statistical, classification, information theory, and clustering-based algorithms. There are many studied that apply anomaly detection algorithms to know normal or abnormal behaviour, flagging irregular behaviour as possibly anomalous. For instance, papers [18, 19] studied anomaly detection algorithms in the context of CPS security. Also, these algorithms are much applicable for other areas like aviation [20, 21], network infrastructure in smart cities [22], smart cities [23], fault detection and tolerant control of nonlinear dynamic system or chaotic systems [24–27] , electric power and smart grid systems [28, 29], web layer cloud [30], and so on.

This study only focusses on machine learning-based anomaly detection algorithms for the CHRIST Osmotron water purification system in which works in pharmaceutical drug manufacture. In the next section, we propose our framework for this aim. Evaluation of the proposed framework comes after.

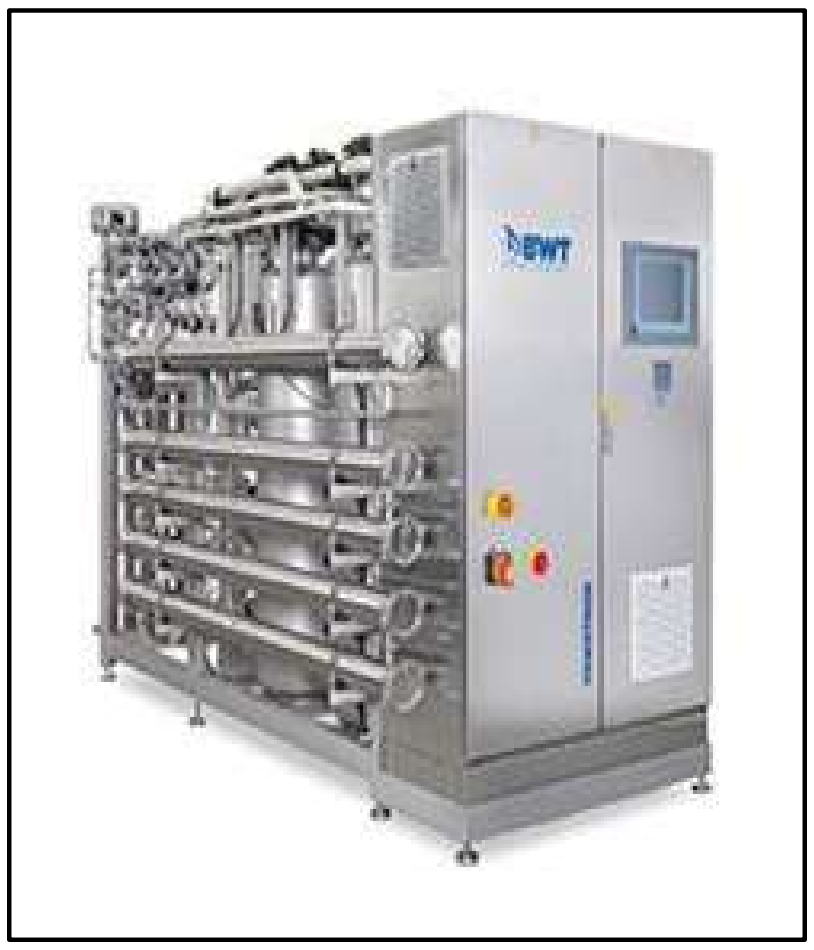

(a) The CHRIST Osmotron system.

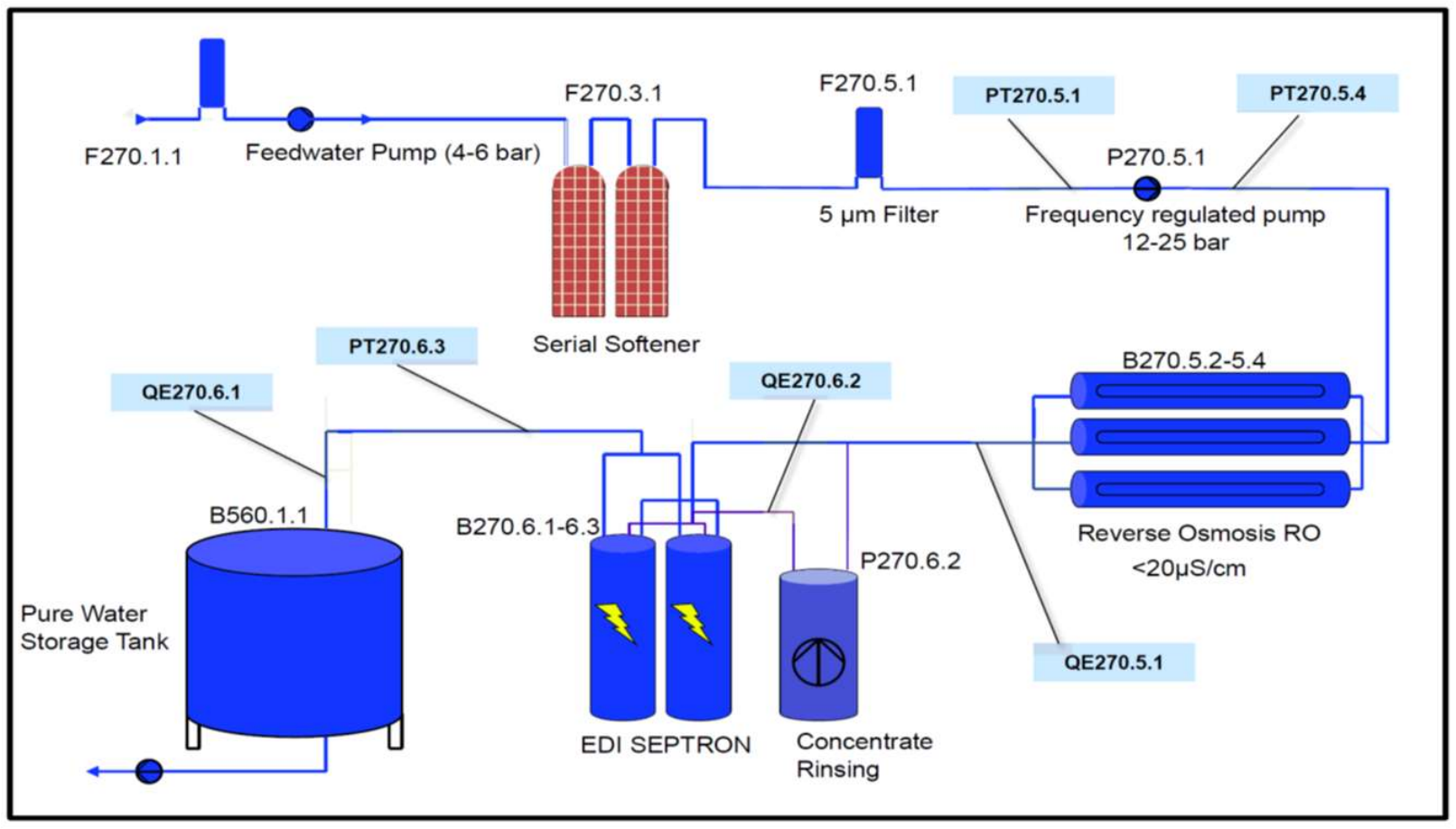

(b) Internals of the Osmotron and the placement of sensors.

Fig. 1. Osmotron® water purification system of CHRIST along with the schematic of its internal components. QE and PT annotations show sensor places.

## III. DESCRIPTION OF CHRIST OSMOTRON® WATER PURIFICATION SYSTEM

The aim of any water purification system is preventing microbial growth and removing particles, organic and inorganic substances, dissolved gases, and microbes. Every phase of production from pretreatment to control of the quality of water, storage and distribution is a major concern in the pharmaceutical industry. In general, there are different processes of water purification according to the United States Pharmacopoeia (USP) and European Pharmacopoeia (EP) requirements including: distillation, reverse osmosis (RO), deionization, ultrafiltration or a combination of them. CHRIST Osmotron® system is purified water by using reverse osmosis and deionization processes and consists of three units which are osmosis, cold and hot loops. The main process of water purification occurs in osmosis unit, and cold and hot loops are used for storage and distribution purified water (PW) and water for injection (WFI), respectively. Briefly, osmosis unit reduces the potable water conductivity from about 700 $\mu m/cm$ to less 1 $\mu m/cm$ by passing water through the softening, membranes and electro-deionize column. The purified water in the outlet of this unit flows to the cold and hot loops.

Scheme of the Osmotron unit of CHRIST Osmotron® water purification system is shown in Fig. 1. Potable water first passes through a filter (F270.1.1) to remove any particles that may be in the water. The filter has 100 $\mu m$ rating. In the next part, a water softener operates on ion-exchange principle and softens the incoming water. It consists of two filters (F270.2.1-3.1) connected in series manner. Calcium and magnesium are replaced by sodium. When the filters become saturated with hardness agents, they have to be regenerated with the aid of sodium chloride (common salt). The softened water is then fed to the reverse osmosis. The reverse osmosis unit (B270.5.2-5.4) serves to demineralize the soft water and operates on the principle of reverse osmosis. The soft water first passes through a fine filter (F270.5.1) to remove any particles that may still in the water. The filter has 5$\mu m$ rating. Then the filtrated water is pumped at high pressure generated by the frequency-regulated pump (P270.5.1) into the permeators. Some of this water passes through the membrane, leaving dissolved substances behind, and leaves the unit as the so-called permeate. The permeate from the reverse osmosis unit enters the EDI SEPTRON modules (B270.6.1-6.3). Each of these consists of two chambers, separated by a special membrane. The pure-water chamber is filled with a special ion-exchanger resin. As the permeate flows through the pure-water chamber, almost all of the ions remaining in it are removed. The ions removed from the permeate are collected in the concentrate chamber, from which they are removed with the rinsing water circulated by the frequency-regulated pump (P270.6.2). The resulting dilute is flowed through the piping to pure water storage tank (B560.1.1) to the consumers.

In the next subsection, two artificial intelligence-based strategies are developed for anomaly detection in CHRIST Osmotron® water purification system.

## IV. ANOMALY DETECTION BY USING ARTIFICIAL INTELLIGENCE

Accuracy in detection and reaction timeliness are important issues in anomaly detection. Reaction time is critical to prevent process failures and cascaded damages in dependable and controllable things. Early detections increase the chance to maintain the system performance. We propose two specific approaches, both based on machine learning techniques, to make an anomaly or fault detection system in CHRIST Osmotron water purification systems. The first one is based on supervised data driven methods that make detection of specific faults using classification methods possible. The second approach is generic and model-based with which any fault can be detected.

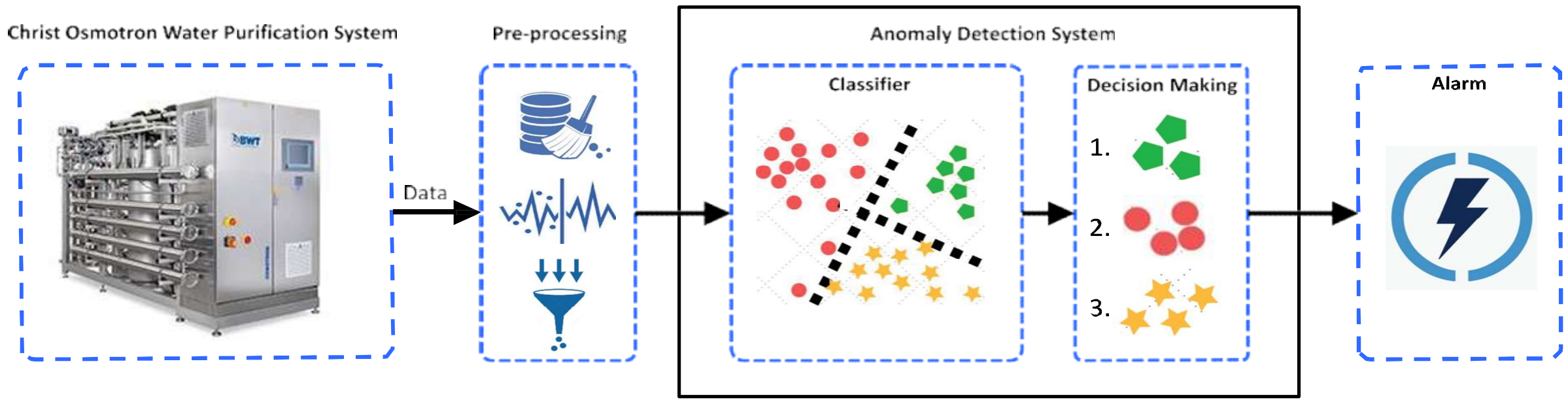

Fig. 2. The first approach for anomaly/fault detection. This approach requires that the abnormal/faulty classes to be known beforehand.

### *A. Approach #1: Data-Driven Anomaly/Fault Detection*

Supervised classification is a machine learning-based technique which is popularly used for detection of previously-seen errors. This is the same technique that is so called signature-based detection in the cyber security domain [29].

In this approach, a labelled training set of normal and faulty samples are collected first and then, the data are fed to a classifier to find a proper boundary that best separates the samples. If there are multiple faults or anomalies of known kind, this method can well detect and separate them. However, similar to any signature-based scheme, the drawback is that if there is a previously unseen fault or anomaly, the detector might not classify it correctly, or even mistake it for normal sample.

Fig. 2 depicts the first approach for anomaly detection in CHRIST Osmotron system. All the steps of this approach except pre-training with labelled samples have been illustrated in this figure. In the results section, we will show how the real-world data collected from a machine before and after overhauling are used to train classifiers like support vector machines (SVM) or artificial neural networks.

### *B. Approach #2: Model-based Anomaly / Fault Detection*

Signature-based detection of Approach 1 is certainly effective if the anomaly or fault attributes/features are known be- forehand. However, it is almost impossible to collect samples of every possible type of fault in a complex interconnected industrial system. The best practices suggest building a normal model of the system instead, which comes to help when one wants to monitor/study the behavior of the system. This approach is generally referred to as anomaly detection in the domain of cyber security [18, 31].

As mentioned before, we collected real-world data from an Osmotron in normal and faulty conditions. However, in this approach, we put the faulty samples aside and only focus on the ones collected during the system's normal operation.

An identifier, which is an artificial neural network, learns the system dynamics under normal conditions. Then, given the input, the developed normal model is used to generate an output estimate and a residual signal ($Y_{out} - Y_{nn}$) which form the basis of our decision making. Note that $Y_{out}$ and $Y_{nn}$ are the system and normal model outputs, respectively.

In this study, we suggest using fixed as well as adaptive thresholds to detect anomalies based on the residue. During the training/learning phase, the system and the identifier outputs form a residual ($Y_{out} - Y_{nn}$) based on which the anomaly detection system determines the detection threshold. The modelling errors collected from the plant/system during the training interval make the threshold for anomaly detection. To calculate the fixed threshold, it is assumed that the residue is an approximation of Gaussian distribution whose mean ($m$) and standard deviation ($v$) of the the residual signal $N(m, v)$ determine a fixed threshold $T$ as follows:

$$T = m \pm \zeta v \tag{1}$$

where $\zeta$ is a (constant) coefficient.

In practice, because of measurement noise and modelling uncertainty, it is essential to make a larger threshold to avoid false alarms. It leads to reduce sensitivity in fault detection. Therefore, the threshold should be chosen such that It compromises between sensitivity of fault decision and rate of the false alarm. Because of mentioned reason, it is recommended to apply adaptive thresholds for industrial applications. Like previous, it is assumed that the residual is an approximation of the normal distribution. The mean and standard deviation are calculated over the past $n$ samples as follows:

$$m(k) = \frac{1}{n} \sum_{i=k-n}^{k} r(i) \tag{2}$$

$$v(k) = \frac{1}{n-1} \sum_{i=k-n}^{k} (r(i) - m(k))^2 \tag{3}$$

(2)

where $0 < n < k$. The adaptive threshold is obtained as follows:

$$T(k) = m(k) \pm \zeta v(k) \tag{4}$$

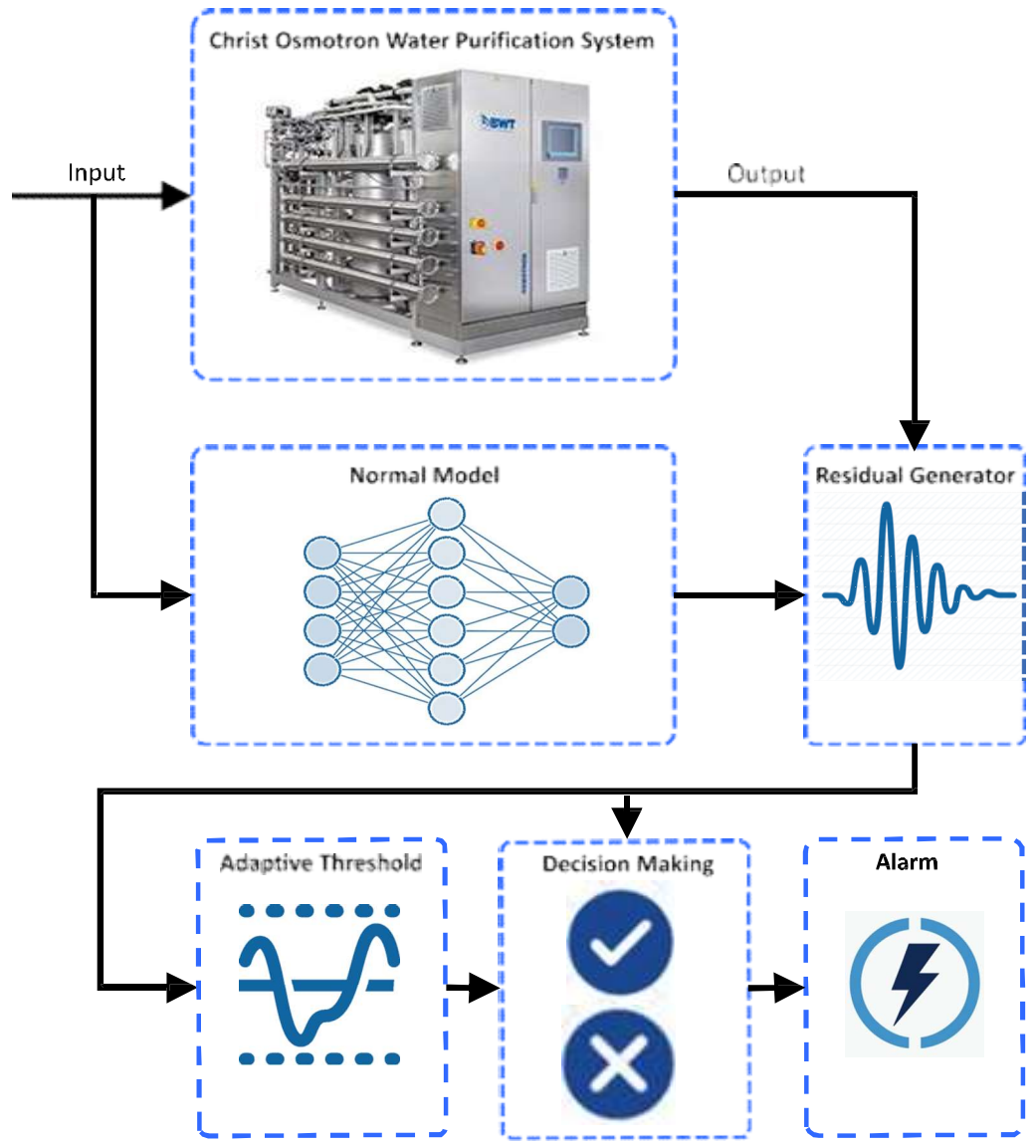

Fig. 3. The second approach for anomaly/fault detection. This approach is more applicable when samples of different system faults are scarce.

Notice that it is important to choose an appropriate length of the time window $n$. If $n$ is selected too small, the threshold quickly adapts to any change in the residual e.g. disturbances, noise or a fault. Besides, if $n$ is too large, the threshold acts similar to a constant one, and leads to reducing the sensitivity of decision making [32].

The second approach can be used in an online manner. The proposed scheme for the second approach of anomaly detection in CHRIST Osmotron system is demonstrated in Fig. (3).

## V. Implementation Results & Discussion

Both approaches for anomaly detection described in the previous section (illustrated in Fig. (2) and Fig. (3)) are applied on CHRIST Osmotron water purification system here.

### *A. Data Gathering*

CHRIST Osmotron® water purification system was installed by its experts in Sina Darou Laboratories Co. in 2006. The device input capacity is 2000 *lit/h*, which under normal conditions should produce 1500 *lit* of purified water with a conductivity of less than 1 *μm/cm*.

Decrease in the output volume of the purified water, as well as increase in the conductivity of the water at the system outlet were two major problems that the device faced at the time of the project launch. Hence, the possible causes and factors of these problems were examined. As mentioned in Section III, membrane filters and EDI ion exchange resins play an important role in the quality and capacity of purified water, so in the first step, membrane filters and then and EDI resins were also replaced during overhaul.

The basic repair and service of the device were included activities such as replacing some O-rings and sealing washers, washing and cleaning the water movement paths, replacing some valves and actuators, calibrating the sensors, replacing some input and output cards of the device controller (PLC), and so on. Moreover, the replacement of the mentioned filters and resins provided the desired performance of the system. Fig. 4a and Fig. 4b show an example of sensors and PLC cards used in this work, respectively.

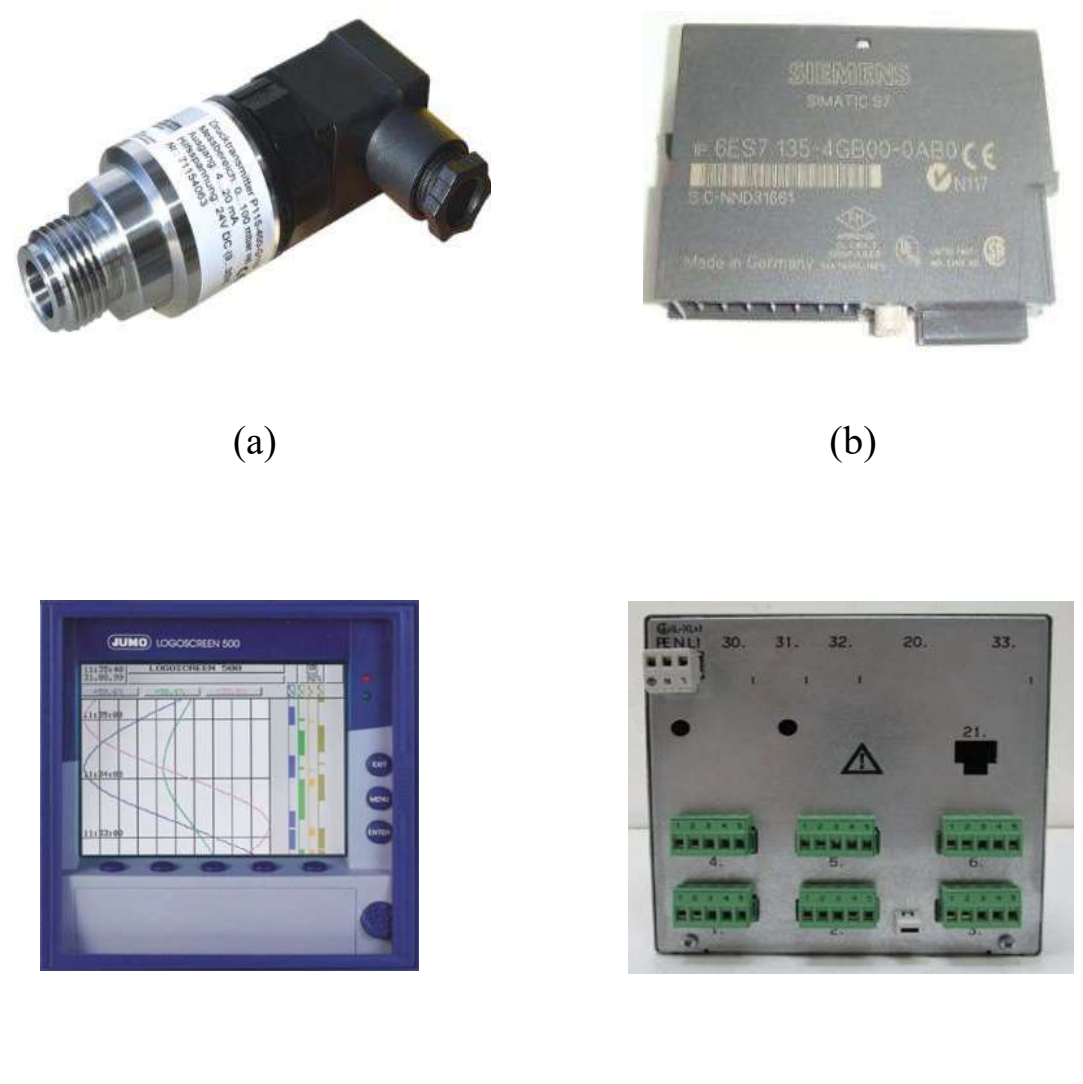

(a) (b) (c) (d)

Fig. 4. (a) A Pressure transducer 4-20 *mA*, 20 to 60 *mbar*, (b) PLC analog output card, (c) Device Screen, (d) Sensor connection terminal.

In this study, six main sensors have been selected and positioned in the Osmotron unit as illustrated in Fig. 1, which are as follows:

- PT270.5.1: Pressure of before frequency regulated pump.
- PT270.5.4: Pressure of after frequency regulated pump.
- QE270.5.1: Conductivity after RO.
- QE270.6.2: Conductivity of concentrate.
- PT270.6.3: Pressure of EDI.
- QE270.6.1: Conductivity of EDI to tank.

Regarding the challenge of data gathering, a data recorder with six-channels and sampling frequency of one minute has been installed in CHRIST Osmotron® system. A Jumo recorder from the LOGOSCREEN 500 series was installed on the Osmotron unit as depicted in Fig. 4c. As shown in Fig. 4d, the recorder can record and store values measured by the mentioned sensors. Then, both normal and faulty data were collected. Faulty data were recorded in the earlier days of the study when the system was suffering from an undesirable performance, and the normal data refers to the data collected after overhaul of the system by manufacturer experts.

A sample of faulty data and normal data (which are collected at intervals of one week and three months after the system repair is depicted in Fig. 5. As observed, faulty data is marked in green and corresponds to one month before the experts repair the system, while the normal data is shown in blue and collected at intervals of one week and three months after repairing the system.

Notice that sampling is done in a minute frequency because of the system variations and tolerance. After gathering data, information were transferred from the recorder to the computer through a memory card.

### *B. Data Preprocessing*

In this study, preprocessing of data consists of three steps; cleansing, normalization, and noise removal.

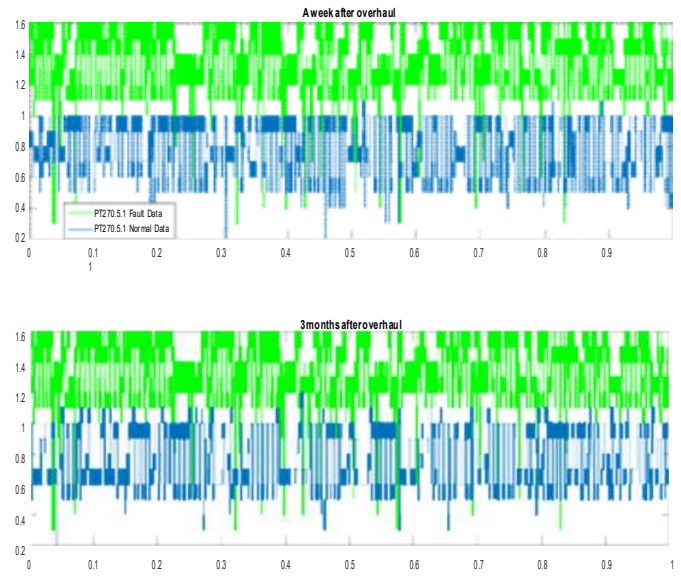

(a) PT270.5.1

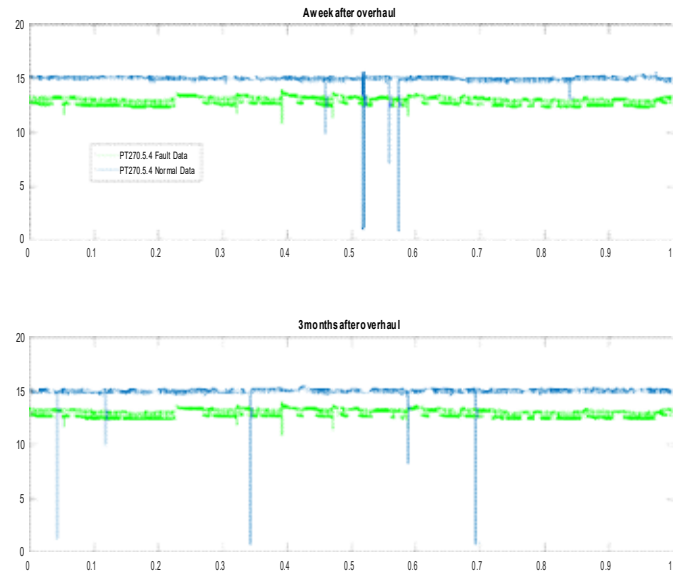

(b) PT270.5.4

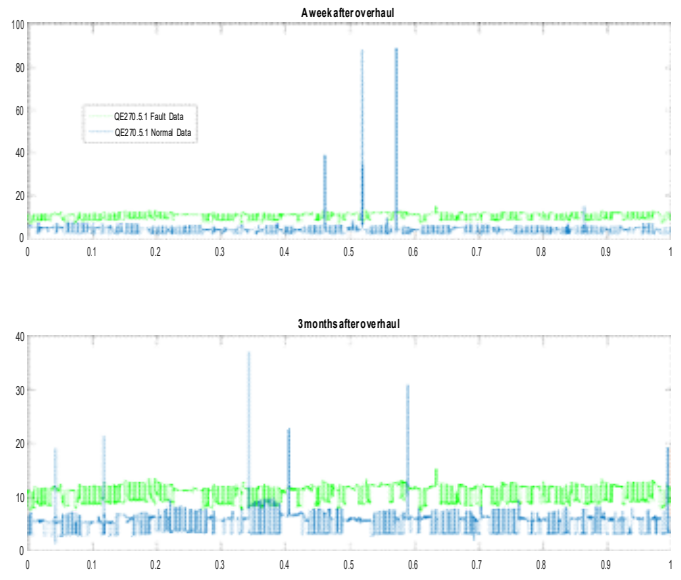

(c) QE270.5.1

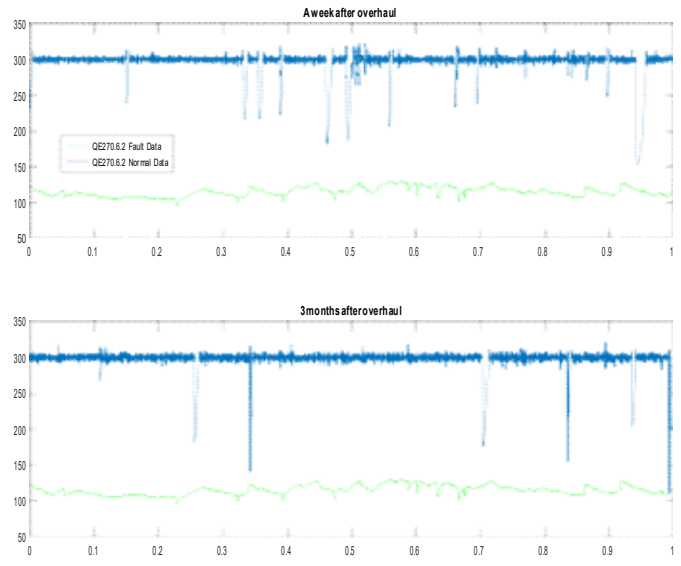

(d) QE270.6.2

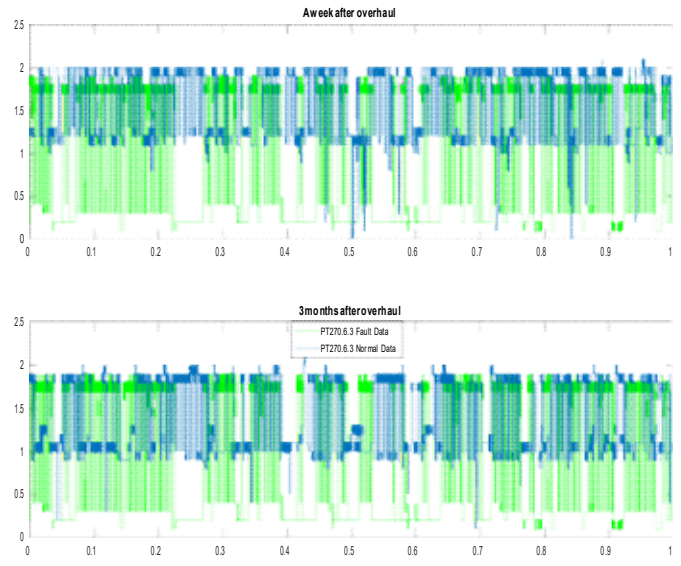

(e) PT270.6.3

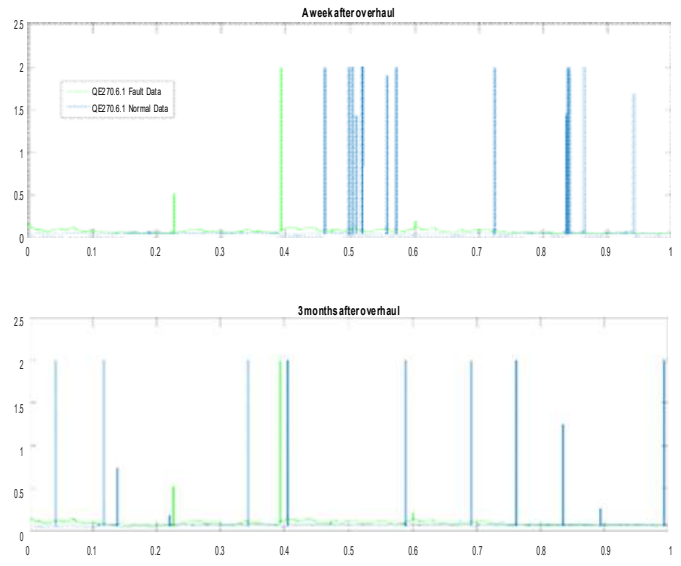

(f) QE270.6.1

Fig. 5. The data collected by six sensors under normal and faulty conditions. The faulty data was collected before overhauling and the normal one was taken twice, once one week after the overhaul and once three months later.

- *Data Cleansing.* Sometimes either the device was shut down for service and filter replacement, or the electricity was cut off, Therefore, some data could be irrelevant or lost. It is important to clean such samples off the dataset.
- *Normalization.* Since the range of variations in the measured data was large, data have been normalized to lie between zero and one.
- *Noise Removal.* To remove and reduce noise effects on the collected data, a low-pass filter, which was Savitzky-Golay filter, was employed. It is a digital smoothing filter.

In the next subsection, the two developed machine learning-based approaches for anomaly detection (illustrated in Fig. 2 and Fig. 3) are applied on CHRIST Osmotron.

## C. Results of Approach #1 (Data-driven)

Whenever we know the faults we are looking for, we can go for supervised signature-based classification. Since we could not deliberately induce different kinds of faults in a sensitive industrial machine like our pharmaceutical water purification system to collect their signatures (or data samples), we merely collected data from the system that had gone faulty mostly because of aging and not being calibrated. Normal samples were collected after overhauling.

We used the classic linear support vector machine (SVM) (as well as neural network and decision tree) as a binary classifier to separate the normal and abnormal samples. Bear in mind that abnormal samples constitute only one kind of fault in this experiment. The six sensors created a set of samples in the 6-dimensional space. Slices of this space in which the input-output relation of important Osmotron components are visible has been depicted in Fig. 6. It is rather easy to find out that the normal operation areas in the majority of these slices are quite distinct from the faulty ones. That is exactly why all of the classic classifiers, including linear SVM, could classify normal and faulty samples with 100% accuracy.

## D. Results of Approach#2 (Model-based)

As mentioned before, since it is almost impossible to have a dataset of all possible faults, we adopt anomaly detection approaches in which only normal operation behavior of the system is learned. This is a common approach in anomaly and intrusion detection of industrial cyber physical systems.

Since we have multiple sensors inside the system, we have access to inputs and outputs of almost every important component. Therefore, rather than identifying the whole system as a black-box based on its inputs and outputs, we can have more granularity and learn each important component behavior separately. For example, let us take the frequency regulated water pump (P270.5.1 in Fig. 1) whose task is to receive water at variable pressures and create an almost constant pressure (around 15 bars) at its output end.

For identification, we took a neural network with two hidden layers. We set the number of neurons at the input layer, the two hidden layers, and the output layer to 5, 22, 20, and 1, respectively. We used 1500 samples in this experiment from which 75% were used in training and the rest were used in the test. Network was trained by using the back propagation method.

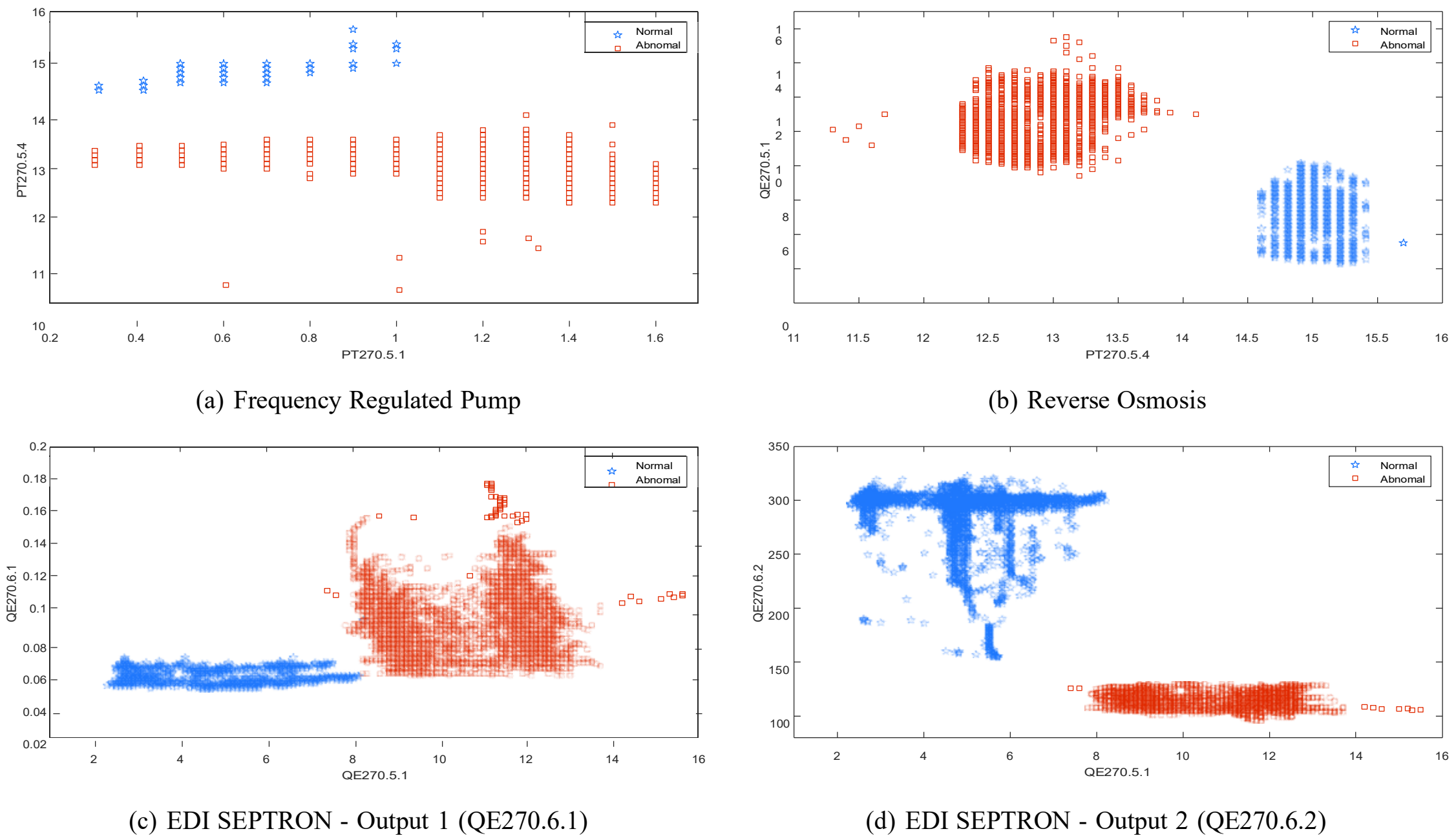

(a) Frequency Regulated Pump (b) Reverse Osmosis

(c) EDI SEPTRON - Output 1 (QE270.6.1) (d) EDI SEPTRON - Output 2 (QE270.6.2)

Fig. 6. Input-Output functional map of (a) Frequency Regulated Pump (b) Reverse Osmosis, (c) EDI SEPTRON - QE270.6.1 , (d) EDI SEPTRON - QE270.6.2.

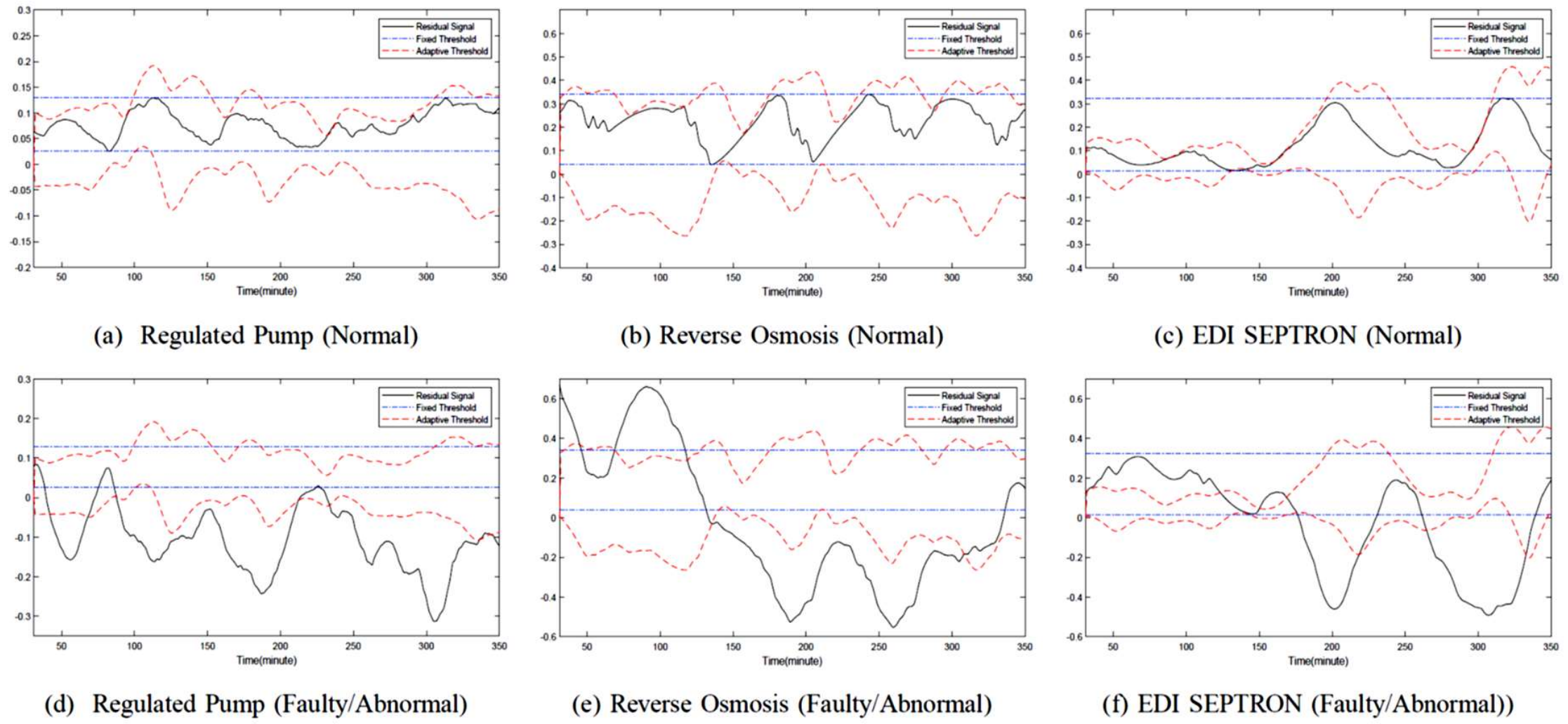

(a) Regulated Pump (Normal) (b) Reverse Osmosis (Normal) (c) EDI SEPTRON (Normal)

(d) Regulated Pump (Faulty/Abnormal) (e) Reverse Osmosis (Faulty/Abnormal) (f) EDI SEPTRON (Faulty/Abnormal))

Fig. 7. Model-based anomaly detection for (a,d) Frequency Regulated Pump (b,e) Reverse Osmosis, and (c,f) EDI SEPTRON. The upper figure shows normal condition in each case while the lower figure demonstrates a faulty/abnormal condition.

Fig. 7 shows the identification residues $(Y_{\text{out}} - Y_{\text{nn}})$ of the pump, reverse osmosis, and EDI SEPTRON when a sequence of test samples (from the normal dataset) is fed to the trained network. It also shows thresholds. For the water pump example, one can see that the maximum estimation error (under normal circumstances) for the pump (Fig. 7a) does not exceed 0.13 bar. If a fixed threshold is to be used for operation anomaly detection, 0.13 can be a good candidate. To see how a faulty system performs, we used the samples of PT270.5.1 and PT270.5.4 before overhauling and plotted the residue found by the same neural network trained by the non-faulty overhauled system. Fig. 7d shows the residue along with the thresholds. It can be seen that a faulty system eventually crosses the threshold and the alarm is raised. While fixed thresholds are handy and easy to find for each system component, they are less sensitive to violations that might happen at certain working points. Therefore, we also tested adaptive thresholds that were updated according to Eq. (4). Fig. 7 plots also include the adaptive threshold boundaries for P270.5.1 (the water pump), reverse osmosis and EDI SEPTRON.

It should be emphasized that unlike the first approach, this detection approach does not rely on previously-seen faulty samples and thus, is more probable to catch unseen anomalies. We can compare the performance of this approach to that of the first one by making a cumulative alarm that is basically the OR'ed alarm signals of the underlying components.

## VI. Conclusion

In this paper, we used two machine learning-based approaches for detecting anomalies or faults in a real-world industrial automation system, i.e. CHRIST Osmotron water purifier. By adding a few sensors and using machine learning techniques, we injected intelligence to this interconnected industrial system and made spotting anomalies and faults possible. A prominent part of this research was collecting real-world data from CHRIST Osmotron before and after overhauling using six sensors. Aging created a kind of fault from which 15000 samples were collected. The same number of samples were collected after overhauling, which represented the system's normal operation attributes. We first used classic supervised classifiers to detect the (one) fault based on the six sensors readings. However, since detection of different or perhaps unknown faults was not easy, we proposed to detect anomalies based on learning the normal behavior of the system and putting the faulty samples aside. Both fixed and adaptive thresholds were tested, and the results showed that while the supervised machine-learning approach is effective, its counterpart has this additional feature that can be used to detect unknown and unseen anomalies.

## Acknowledgement

We are thankful to SinaDarou Laboratories Company, and specifically its CEO, Dr. Mohammad Hasan Vasefi, and the Manager of Engineering and Technical Department, Mr. Ali Naghi Mansubi, for their supports during this study.